\documentclass[article]{revtex4}
\begin{document}
\title{Securing QKD Links in the Full Hilbert Space}
\author{Deborah J. Jackson}
\author{George M. Hockney}
\affiliation{%
Quantum Computing Technologies\\
 Jet Propulsion Laboratory, California Institute of Technology \\
4800 Oak Grove Drive, Pasadena, California 91109-8099 \\
}%
\date{\today}
\begin{abstract}
\begin{center}
\bf Many QKD analyses examine the link security in a subset of the full Hilbert space that is available to describe the system. In reality, information about the photon state can be embedded in correlations between the polarization space and other dimensions of the full Hilbert space in such a way that Eve can determine the polarization of a photon without affecting it.  This paper uses the concept of suitability \cite{Hockney-02} to quantify the available information for Eve to exploit, and and demonstrate how it is possible for Alice and Bob to fool themselves into thinking they have a highly secure link.    
\end{center}

~~~~~

\end{abstract}
\maketitle
\parskip 3pt

~~~~~

\section{Introduction}
{T}{he} earliest treatment of QKD \cite{Bennett-84}  was a theoretical paper that defined a method for distant parties (Alice and Bob) to agree on a random sequence of key bits through the information encoded and transferred between the them.  This approach relies on polarization encoding of photons and is now well known as the BB84 protocol.  Since then, experimenters have used faint laser sources to demonstrate key exchanges via polarization bases\cite{Muller-93} \cite{Breguet-94} \cite{Muller-95},  phase encoded qubits \cite{Bennett-92} \cite{Townsend-93} \cite{Hughes-96}  \cite{Breguet-95}, and frequency encoded qubits \cite{Mazurenko-97} \cite{Sun-95} \cite{Molotkov-98}.  Another important trend has been the use of entangled photon pairs for quantum cryptography.  Thus far, experiments exploiting the polarization entangled photons \cite{Jennewein-00} \cite{Naik-00}   as well as implementations taking advantage of energy-time entangled photons \cite{Franson-89} \cite{Ribordy-01} \cite{Tittel-00}  have been reported.  Although the implementation of entanglement experiments tends to be more complex than implementations using faint laser sources, it is generally believed \cite{Mayers-98}  that entangled photon pairs are more desirable because they prevent unintended use of the extra degrees of freedom in the photon number state. The truth is that the equipment used to build the experiments  is not perfect, so that the advantage of one systemÕs approach over another changes depending on oneÕs choice of implementation hardware.  In other words, under some conditions (e.g. a measurement setup with an extremely low quantum efficiency detector) the faint photon source might have a significant advantage over the SPDC down conversion source, while a different set of  conditions and hardware might yield a significant advantage to the system with the SPDC source.  Indeed, many of the early analyses of proposed QKD systems assumed perfect detectors, sources, beam splitters, and other apparatus, and failed to take into account real world imperfections.  However, as the field matured, imperfections in sources and detectors began to drive the architectures \cite{Lutkenhaus-99}  \cite{Gilbert-00}.  Now the introduction of the suitability parameter gives a systematic set of ground rules for comparing both different hardware implementations of a single approach as well as completely different approaches.  
This paper takes the suitability concept defined in Hockney et. al. \cite{Hockney-02}  and applies it to the quantum key distribution (QKD) application. In that paper, the basic concept of suitability was introduced as a means of helping one compare the usefulness of individual hardware devices (e.g. detectors, sources, etc.) for quantum computing applications.  

 After showing how losses in the channel adversely affect the security of the exchange link in the expected fashion, this paper then uses the suitability to examine how parameters other than the number state and the polarization state can impact the security of quantum key distribution.   
The general 
definition of the suitability, $S_{GT}$, describes, on a scale from 0 to 1, how well a given source gun, G, and a given target system, T, will work together.   The suitability calculation is always performed 
using the requirements of a specific application context. For example, the QKD 
application has two classes of users  (i) those who need to share a secret 
(Alice, Bob, and their associates) and (ii) those who wish to surreptitiously discover 
shared secrets (Eve and her 
associates).   For Alice and Bob, the target T is the apparatus used to extract polarization information.  In addition, Alice and Bob need to compute the suitability for the worst case of information leakage to Eve.  That is, they assume that any photons which they cannot account for have been captured and "read" by Eve. In the example discussed in this paper, the 
suitability factor clearly shows how efficiently information carried by the 
photon is distributed from Alice to Bob and how effectively the information 
shared by Alice and Bob can be kept from Eve.  Furthermore, though Alice and 
Bob only need to work in the polarization subset of the full Hilbert space to 
establish a shared key when using the BB84 protocol, it becomes very clear that they must work in the full 
Hilbert space to properly evaluate the security of their key exchange system.  
The organization of this paper is as follows.  Section II defines the suitability factor for the polarization-only Hilbert space.  Section 
III discusses how Eve can improve her opportunity to cheat, Section IV 
discusses the need for treating the full Hilbert space to minimize Eve's cheating opportunities, and Section V draws 
conclusions from what we have learned in this treatment. 

\section{Polarization Hilbert Space}
This section reviews concepts that were introduced in reference \cite{Hockney-02}.   In this discussion, a measurement system will consist of the following components:
a source, a transmission channel, one or more analyzers, and a detector.  The arrangement of these
 components is determined by the specific objectives of the parties involved in 
the measurement. The suitability is defined quantum mechanically in terms of 
the density matrix, 

\begin{equation}
\rho_{T}=\frac{1}{N_{T}}\sum_{i=1}|{\Psi_{i}}\rangle\langle{\Psi_i}|
\end{equation}

for the $N_{T}$ possible target states suitable for detection by the detector; 
and the gun density matrix, $\rho_{G}$,  the probability distribution of all possible 
output states that the source gun can produce.  These are then used to calculate 
the suitability, which is the probability that the output of the source gun will be usable by the detector.  For example, if the source is a pure state and the target is a pure state, $S_{GT}$  reduces to the wave function overlap between the source and the detector states.

\begin{equation}
S_{GT}=\frac{F_{GT}}{F_{TT}}=\frac{Tr(\rho_{GT})}{Tr(\rho_{TT})},
\end{equation}

where the quantity,  $F_{ij}=Tr(\rho_{i}\rho_{j})$ for any two normalized density matrices, 
$\rho_{i}$ and $\rho_{j}$.  
Note that the density matrix of a pure state satisfies the condition  
$Tr(\rho^{2})=1$, which is physically interpreted as the probability that two consecutive shots from the gun are the same state.  Whereas the purpose of defining the suitability in reference \cite{Hockney-02} was to permit one to rank the usefulness of the individual devices that are being developed by researchers for quantum computing applications, this paper shows how it can be applied to the comparison of different physical implementations of integrated hardware systems.  Prior results have always been based on ad hoc probability calculations that define an idealized measurement space which is usually a subset of the real physical measurement space.   Let us take
the example of a QKD network consisting of two nodes belonging to Alice
and Bob, who have chosen to implement a simple two-state variation of the BB84 protocol \cite{Bennett-84}.  Because Alice and Bob have agreed on the system design beforehand, they know the protocol and everything about their own measurement system, so it is easy for them to calculate the suitability of their setup for key construction.  In this example the channel is assumed to be lossless, and the detectorÕs quantum efficiency is assumed to be $\eta_{B} = 1$. This specifically does not depend on the entire Hilbert space, but only on the sub-space involving polarization in a well defined time slot.  If one constructs a truth table of all combinations of Alice's and BobÕs polarization settings, there are four possible measurement configurations. The suitability can be used to determine the probability that the detector will fire for each configuration.

This is obtained by starting with the density matrix of the gun-state, $\rho_{G}$,  during the time slot, $t_{0}$, allowing any changes induced by passage through the channel to act upon the gun-state density matrix, and then taking the product of  the target-state density matrix, $\rho_{T}$, in  the associated timeslot, $t_{0}+L_{AB}/c$. Note that since BobÕs polarizer in each configuration defines only one target-state, $N_{T}=1$ in equation 1.   Here $L_{AB}$ is defined as the length of the fiber separating AliceÕs  source from BobÕs detector and $c$ is the speed of light.  Assuming that there are no loss mechanisms between the two nodes, BobÕs 
detector will fire as long as one of the detection states,  $|{\Psi_{i}}\rangle$,  
overlaps with the 
gunÕs single-photon state. In this system, the combined orientations of AliceÕs 
polarization rotator and BobÕs polarization analyzer define the relevant information 
encoded in the channel and the suitability is calculated by following three simple 
rules:

\hspace{0.25 in}    $\bullet$ Conditions imposed by AliceÕs preparation are incorporated into the gun density matrix, $\rho_{G}$.

\hspace{0.25 in}	$\bullet$ Conditions imposed by BobÕs analyzer measurements are incorporated into the target density matrix, $\rho_{T}$.

\hspace{0.25in}	$\bullet$ Channel losses are incorporated into the source state density matrix, $\rho_{G}$.

The density matrices for AliceÕs polarization states are derived from equation [1] using the
 following definitions for AliceÕs source states \[|{\Pi_{0}}\rangle_{A}\ = |{H}\rangle\]  and \[ |{\Pi_{1}}\rangle_{A} = (|{H}\rangle-|{V}\rangle)/ {\sqrt{2}},\].   Therefore the gun density matrices are:

\begin{equation}
\rho_{A,0}= \left( \begin{array}{cc} 1 & 0 \\ 0 & 0 \end{array}\right); 
	\rho_{A,1}=\frac{1}{2} \left( \begin{array}{cc} 1 & -1 \\ -1 & 1 \end{array}\right),  
\end{equation}

while BobÕs analyzer, which defines the target states \[|{\Pi_{0}}\rangle_{B } = (|{H}\rangle+|{V}\rangle)/ {\sqrt{2}}\]  and \[|{\Pi_{1}}\rangle_{B} = |{V}\rangle,\] yields the target 
density matrices:

\begin{equation}
\rho_{B,0}=\frac{1}{2} \left( \begin{array}{cc} 1 & 1 \\ 1 & 1 \end{array}\right);  		\rho_{B,1}=\left( \begin{array}{cc} 0 & 0 \\ 0 & 1 \end{array}\right). 
\end{equation}

In a lossless system with perfect detectors, the suitability for key construction is
\begin{equation}
S_{AB_{ij}}=F_{AB_{ij}}=Tr(\rho_{A,i}\rho_{B,j}) = \frac{1}{2}\delta_{ij}, 
\end{equation}

The factor of 1/2, which explicitly reflects the probability that Bob receives a usable photon, is exactly what one expects when both he and Alice select the same bit-value (i.e. $i = j$), ; Bob will gain a bit of shared key half of the time.  It tells one how efficiently information is transferred via  the projective measurements between Alice and Bob. 
	
\section{Opportunities for Eve}
 We can also rate on a scale from 0 to 1, the suitability of Alice's source for the eavesdropper, Eve. The classical treatment of QKD security depends on the non-clonability feature of qubits. But within the classic QKD scenario, there is an implicit assumption that the polarization Hilbert space uniquely describes everything one needs to know about the key distribution system.  The traditional treatments of QKD security tend to focus on the ability of an eaves-dropper to carry out undetected polarization measurements and assumes that the polarization state is the only way that the photon carries its encoded information.  More specifically, it assumes that any variation of the measurement apparatus that Eve can bring to bear in order  to extract information about the photon is limited to manipulations or measurements within the polarization Hilbert space. In fact, the real Hilbert space is much larger, including descriptions of timing, frequency, phase, and number state for the photon.  This means that Eve can take advantage of the larger Hilbert space and cheat, especially  if the information encoded wave functions  do not overlap in the parts of the Hilbert space that are not part of Alice and Bob's protocol definitions. In the discussion that follows, we first consider what happens in the ordinary polarization encoding case, and then we add in the other dimensions of Hilbert space to show how it opens up opportunities that Eve can exploit to her advantage.

The problem from Bob's and AliceÕs perspectives is that they really don't know what protocol Eve will chose in trying to recover information on the key; remember that Eve is allowed to do anything.  Thus if Bob and Alice try to calculate Eve's suitability, they must focus on defining the maximum amount of information that Eve could extract from the channel (without changing the photon statistics for their measurements).

\begin{figure}
[top] figure1.eps
\caption{Schematic of projective measurement setup between Alice and Bob for the purpose of agreeing on a secret key.  The "cloud" around Eve represent unknowns in the measurement that might allow Eve to extract information about the key.}
\end{figure}

The cloud in Figure ~1 represents Bob's and AliceÕs uncertainty in terms of EveÕs protocol and measurement methods. The conservative approach for  Alice and Bob would be to assume that the cloud has ideal properties for the eavesdropper in that it (1) passes all single photons to Bob that are emitted from AliceÕs photon gun, and (2) when more than one photon is incident during the same time slot, the cloud diverts the extra photons to Eve with with 100\% probability.  Thus in the situation where one has a true single-photon source, $\rho_{E,I} = 0$, because no photons are diverted to Eve; hence the suitability for Eve to recover the key sequence is given by  \[S_{AE_{ij}} = F_{AE_{ij}} =  Tr( \rho_{A,i} \rho_{E,j})=0.\]  

When Alice has an imperfect source, such as the attenuated pulsed lasers actually used in QKD experiments, the wave function for the pseudo-single photon sources spans a larger Hilbert space, which must be taken into account.  These are  best represented as weak coherent states of the radiation field. This means that to be completely correct, the initial wavefunction must describe both the polarization state of the photon and the number state.

\begin{equation}
|\psi_{i}\rangle=\sum_{n} c_{ni}|n\rangle|\Pi_{i}\rangle. 
\end{equation}

 For example, a weak source would have
\[c_{ni} = e^{\frac{-|\alpha|^{2}}{2}}\frac{\alpha^{n}}{\sqrt{n!}}\]
The sum over $n$ is over the full Hilbert space of number states, $|\Pi_{i}\rangle$  represents the polarization states, the sum over $i$  is the sum over the polarization Hilbert space, and $\alpha$ is the expectation value of the number operator.   The new density matrix for this source after interaction with the environment of the channel (i.e. AliceÕs rotation, BobÕs analyzer, and any losses incurred in the channel) is:

\begin{equation}
S_{AB_{ij}}=F_{AB_{ij}}=Tr([e^{-|\alpha|^{2}}\sum_{m=0}^{\infty}\frac{\alpha^{2m}}{m!}||{m}\rangle\langle{m}|]\rho_{A,i}[e^{-|\alpha|^{2}}\sum_{n=1}^{\infty}\frac{\alpha^{2n}}{n!}|{n}\rangle\langle{n}|]\rho_{B,j}). \end{equation}

Gisin, et. al. \cite{Gisin-02} summarize the body of literature that explores how much information Eve can potentially exploit in the presence of more than one photon in the channel.  With a heavily attenuated source, (which produces the vacuum state,  $|{0}\rangle$, on average, occasionally a single-photon, $|{1}\rangle$, and more rarely multiple photons,  $|{n}\rangle$, for $n \geq 2$) BobÕs suitability to detect the generated key is reduced by the probability that zero photons are generated in a given time slot.  That is, as long as one or more photons are created by Alice, Bob has a finite probability of being able to observe the photon.  In addition, with such a source, things also become very interesting for Eve because every now and then, $n \geq 2$ so that EveÕs suitability takes on finite values.  Now Bob and EveÕs respective suitabilities for any given outcome are:

\begin{eqnarray}
{S_{AB_{ij}}=F_{AB_{ij}}=Tr([e^{-|\alpha|^{2}}\sum_{m=0}^{\infty}\frac{\alpha^{2m}}{m!}|{m}\rangle\langle{m}|]\rho_{A,i}[e^{-|\alpha|^{2}}\sum_{n=1}^{\infty}\frac{\alpha^{2n}}{n!}|{n}\rangle\langle{n}|]\rho_{B,j})} \\
=\frac{(1-e^{-|\alpha|^{2}})\delta_{ij}}{2}. 
\end{eqnarray}

\begin{eqnarray}
{S_{AE_{ij}}=F_{AE_{ij}}=Tr([e^{-|\alpha|^{2}}\sum_{m=0}^{\infty}\frac{\alpha^{2m}}{m!}||{m}\rangle\langle{m}|]\rho_{A,i}[e^{-|\alpha|^{2}}\sum_{n=2}^{\infty}\frac{\alpha^{2n}}{n!}|{n}\rangle\langle{n}|]\rho_{E,j})} \\
=\frac{(1-e^{-|\alpha|^{2}}-\alpha^{2}e^{-|\alpha|^{2}})\delta_{ij}}{2}. 
\end{eqnarray}

Because $S_{AB}$ is defined only for those states where $n \geq 1$,   $Tr(\rho_{A}) = 1$.  Likewise, $S_{AE}$ is defined only for those states  when  $n \geq 2$,  hence $Tr(\rho_{E}) = 1$.    Assuming that Eve is performing the same measurements as Bob in her attempts to extract a key, the ratio of the {sum of all Eve's measurement configurations to Bob's measurement configurations is}
 
\begin{equation}
\frac{S_{AE}}{S_{AB}}=\frac{\sum_{ij}S_{AE_{ij}}}{\sum_{i'j'}S_{AB_{i'j'}}}=\Gamma,    
\end{equation}

which determines the percentage of the time that Eve extracts information on the exchanged key bits with her measurements.  Here $\Gamma=(1-e^{-|\alpha|^{2}}-\alpha^{2}e^{-|\alpha|^{2}})/ (1-e^{-|\alpha|^{2}})$. For a coherent source in the single photon limit, $e^{-|\alpha|^{2}} \simeq1-|\alpha|^{2}$ 

\begin{equation}
\frac{S_{AE}}{S_{AB}}=\Gamma \simeq |\alpha|^{2}      \label{eq:privacy}
\end{equation}

Notice that EveÕs suitability has thus far always been calculated from the perspective of what Alice and Bob know about Eve.   If Eve were to calculate the suitability herself the outcome would be completely different because she knows everything about the limitations of her own measurement setup and hence can calculate her suitability more exactly than Bob and Alice can.  Furthermore, the emphasis of her suitability calculation differs because what she really wants to know is how to the maximize the probability of recovering key information while minimizing the chances that either Alice or Bob detect her presence on the channel. Since there are some aspects of Alice's and BobÕs measurement system of which Eve is unaware, her uncertainty about these details will be reflected as a larger uncertainty in EveÕs confidence that Alice and Bob have not detected her presence on the channel.

\section{Full Hilbert Space}

However, the full Hilbert space of the channel is much larger.  It contains timing information and, if it is not in a single mode fiber, mode information.  In this section we expand the wave function to include the other degrees of freedom in the photon state that are not intentionally encoded as part of the key exchange protocol.  The model thus far has concentrated on the polarization subspace and the number states defined by the coherent states of the radiation field. However, the Hilbert space that can be exploited for encoding information onto a single photon is still larger than the one defined by the polarization basis and the number states.  Other properties which are exploitable by Eve are the photon arrival time  (which changes its position in the time slot), the photon wave number, and the photon energy\cite{Franson-89}.      Hence the full wavefunction is of the form:

\begin{equation}
|{\Psi_{i}(t_{i})}\rangle = \{ \int_{t } \int_{\vec{k}} \sum_{n}^{\infty} c_{ni}(t_{i},\vec{k}) |{n}\rangle  |{\Omega(t - t_{i})}\rangle |{K(\vec{k})}\rangle dt d\vec{k}\}|{\Pi_{i}}\rangle,  				 \label{eq:hilbert}
\end{equation}

where  $|{\Omega(t-t_{i})}\rangle$  is the wavefunction describing the photonÕs temporal behavior at time $t$, and  $|{K(\vec{k})}\rangle$ is the normalized wavefunction for the photon wave vectors.  For example, recent findings that single photons carry orbital angular momentum \cite{Allen-92}, imply that  $|{\Pi_{i}}\rangle$ should include both the spin angular momentum due to polarization and the orbital angular momentum associated with the momentum vector.  Together, they would both span a larger Hilbert space than the $i=2$ due to the polarization states alone.     

Because of these other dimensions Eve can potentially extract  key information without performing a direct polarization analysis on the photon.  In the example given above, Alice and Bob only perform projective measurements on the polarization state because that is all they need to know to complete their key exchange.  Thus their suitability for completing the key exchange is unchanged from equation 10, despite the more complex wavefunction.  On the other hand, suppose that Alice'Õs source had some inadvertent, but measurable bias built into the photon time of arrival (as shown in Figure ~2) and that this timing bias was a direct function of the selected polarization state. (This could happen since birefringence in the half wave plates used for polarization rotation clearly modify the temporal behavior of the wave packets by retarding one polarization more than the other.) If, in the temporal Hilbert space, the polarization states are distinguishable and if Eve has modified her detection apparatus to measure the arrival time of the photon in fine enough time increments, she can infer information about the polarization state.  Furthermore, the use of a high bandwidth quantum non-demolition \cite{Kok-02} apparatus would enable her to infer this information without directly detecting the photon.  With such a detector, EveÕs suitability is based on the extent to which the wavefunctions for the different polarization states are distinguishable during EveÕs measurements.  Eve measures
\begin{equation}
I(t)=  | |{\Psi_{i}(t-t_{i})}\rangle + |{\Psi_{j}(t-t_{j})}\rangle \|^{2},  		
\end{equation}

which will have an interference term when the polarization states are completely  indistinguishable in time domain measurements if $|t_{i} - t_{j}| = 0$.  Defining $\Delta T$ as the temporal size of the wavefunction, and $\delta T$ as the temporal resolution of Eve's measurement apparatus, one also sees that the polarization states are completely distinguishable in the time domain when $|t_{i} - t_{j}| >> \Delta T$ and $\Delta T > \delta T$.   Notice that in all or our earlier treatments, the information, conveyed by the polarization state, was extracted via a projective measurement into a specific state.  However, in the temporal domain, Eve can setup to measure everything that comes her way by sorting them into time bins.  The only requirement is that her temporal measurement resolution must be smaller than the time separation between the two states.   As a consequence, the best test for determining distinguishability is to look for evidence of the interference cross terms.

\begin{figure}
[top] 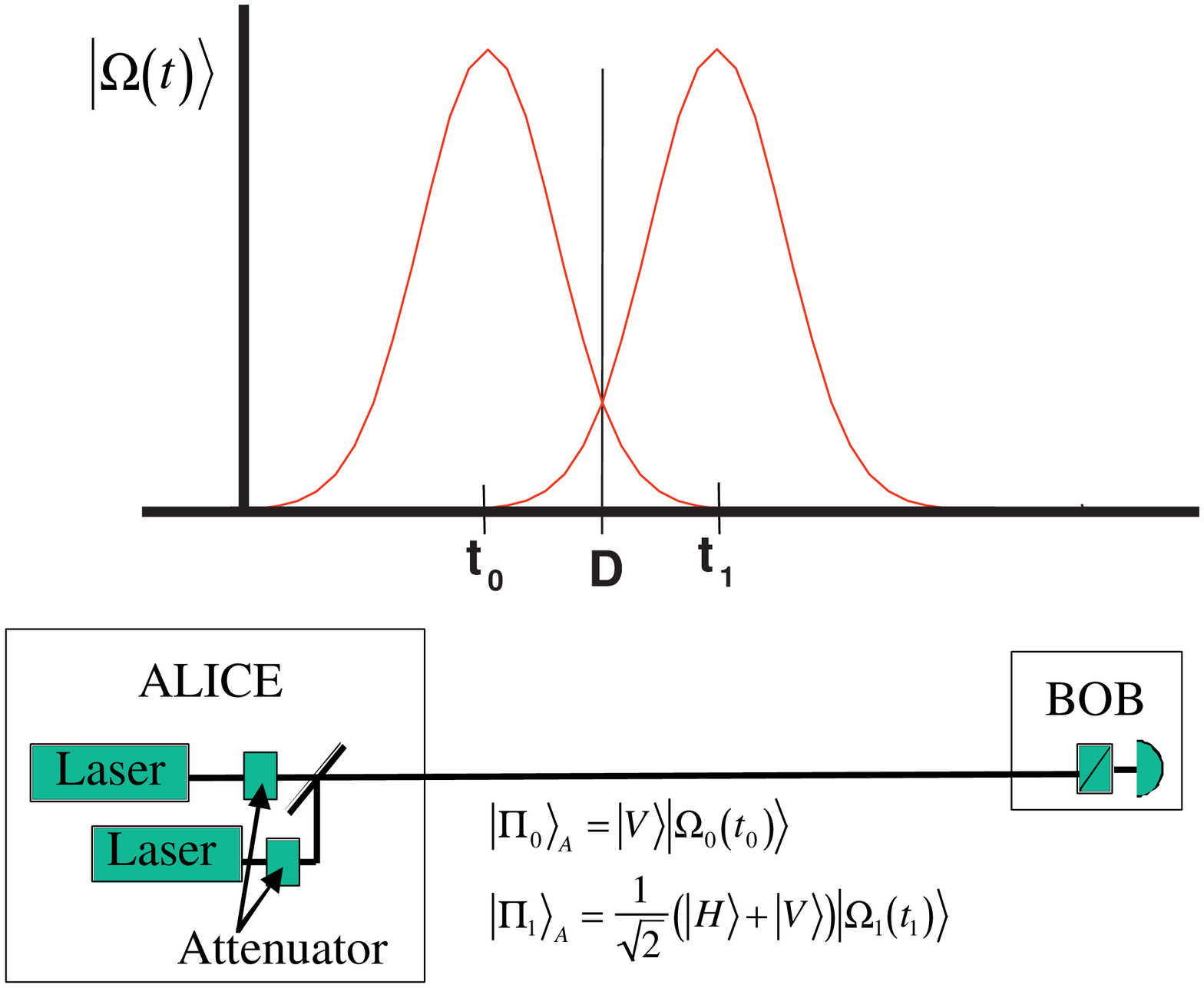
\caption{By chosing to make measurements in a different Hilbert space from the one used by Alice and Bob for the key exchange, Eve may be able to extract some or all of the key if Alice and Bob are sloppy in setting up their projective measurement hardware.}
\end{figure}
To check the suitability for Eve, Alice and Bob must calculate the extent to which the two different polarization states are indistinguishable in the other dimensions of the full Hilbert space.  Clearly, if EveÕs measurement instruments can resolve the  non-overlapping wavefunctions of  the different photon states in the time or in the frequency subspaces, Eve would be able to extract more information from the channel than Bob could with his polarization analyzers. It thus behooves Eve to design and build an apparatus which is capable of resolving the non-overlap portions of the wave function in a dimension orthogonal to the one chosen for the key exchange protocol.  If Eve is successful in the temporal domain, the respective suitabilities are 

\begin{equation}
S_{AB_{ij}}=\frac{(1-e^{-|\alpha|^{2}})\delta_{ij}}{2}.
\end{equation}

\begin{eqnarray}
S_{AE_{ij}} = F_{AE_{ij}} =Tr(\rho_{A,i}[1-\frac{1}{N_{T}}\{|{\Psi_{i}(t_{i})}\rangle\langle{\Psi_{j}(t_{j})}|+ |{\Psi_{j}(t_{j})}\rangle\langle{\Psi_{i}(t_{i})}|\}]) \\
=\frac{1}{2}\delta_{ij}[1-\frac{1}{N_{T}}\{|{\Psi_{i}(t_{i})}\rangle\langle{\Psi_{j}(t_{j})}| +  |{\Psi_{j}(t_{j})}\rangle\langle{\Psi_{i}(t_{i})}|\}].
\end{eqnarray}

where $\Psi(t_{i})$ and $\Psi(t_{i})$ denote the temporal evolution of the encoded source states.  But Eve is not limited to the temporal Hilbert space. She can also devise hardware to distinguish other properties of the photon such as frequency, momentum vector, etc.  such that  the percentage of Bob's  bits that Eve recovers is
 
\begin{equation}
\frac{S_{AE}}{S_{AB}}=\frac{\sum_{ij}S_{AE_{ij}}}{\sum_{i'j'}S_{AB_{i'j'}}}=\frac{[1-\frac{1}{N_{T}}\{|{\Psi_{i}(t_{i},\vec{k}_{i})}\rangle \langle{\Psi_{j}(t_{j},\vec{k}_{j})}| + |{\Psi_{j}(t_{j},\vec{k}_{j})}\rangle \langle{\Psi_{i}(t_{i},\vec{k}_{i})}|\}]}{(1-e^{-|\alpha|^{2}})}.    		 \label{eq:general}
\end{equation}

Here, $N_{T } = 2$ because Alice and Bob encoded a total of two different target states. This calculation is based on Alice and Bob's understanding of what information Eve could capture when the wave functions of their encoded states have imperfect overlap.   To state it more generally, the overlap of the wavefunctions for AliceÕs two polarization states determine how much information Eve can extract from the channel through measurements in a subspaces orthogonal to the polarization subspace which was selected for encoding the key.  If the overlap is perfect, there is sufficient ambiguity about the photon state that Eve gets no information, but if there is absolutely no overlap between the two wavefunctions and Eve can resolve it; then Eve gets practically everything. This means that Eve knows exactly which of her measurements in the orthogonal subspace corresponds to AliceÕs 0-bit and 1-bit polarization values, which means there is no security for the key exchange because Eve's measurements allow her to read out the polarization encoding imparted to each photon.   Furthermore, when Eve's suitability for eavesdropping is greater than or equal to Alice's and BobÕs suitability for key construction, the privacy amplification protocol developed by Bennett et. al. \cite{Bennett-95} breaks down.  In essence, more bits must be removed to insure security than exist in the exchanged key.  Consequently, it is critical for Alice and Bob to include in the secrecy analysis of their exchange protocol, all of the Hilbert subspaces beyond the one in which the qubit information is intentionally encoded.  The analysis is performed in these other subspaces in order to define the criteria for closing potential information extraction loopholes that Eve could exploit.  The loopholes are closed  by imposing additional performance specifications on the measurable effects of the source, analyzers, transmission medium, etc. on the qubit wavefunction for these additional subspaces.   

\section{Conclusions}
Suitability provides a systematic method of determining both the key exchange efficiency of networks and the unintentional weakness in a specific hardware implementation of networks to eavesdropping. In determining the security of a key exchange channel, we have found that it is important to make a mathematical model that includes all of the physical variables of the system.  Writing down the entire wavefunction that spans the full Hilbert space would allow one to quickly identify alternative measurements that Eve might exploit to extract information about the exchanged key.   As shown here, systematic application of suitability criterion over the full measurement Hilbert space, when designing key exchange networks should help improve the security of the final product.  If we assume that Eve can only access the channel, there is still the question of whether Alice's description of the channel is complete.  However, even if Equation~\ref{eq:hilbert}  does not quite span the full Hilbert space of the channel, one sees from Equation~\ref{eq:general}  that it is possible to verify that  the wavefunctions have good overlap across the full Hilbert space  if one observes, through measurement, that the interference terms are present.  The Hong, Ou, Mandel experiment \cite{Hong-87} both describes and demonstrates a prototype setup for determining wavefunction overlap across the full Hilbert space.  
\section{Acknowledgement}
The research described in this (publication or paper) was carried out at the Jet Propulsion Laboratory, California Institute of Technology, under a contract with the National Aeronautics and Space Administration.

\end{document}